\documentstyle[preprint,tighten,aps,epsfig,colordvi,psfig]{revtex}

\def\spose#1{\hbox to 0pt{#1\hss}}
\def\ltapprox{\mathrel{\spose{\lower 3pt\hbox{$\mathchar"218$}}
 \raise 2.0pt\hbox{$\mathchar"13C$}}}
\begin{document}
\draft
\vskip 2cm

\title{Improved Perturbation Theory for Improved Lattice Actions}

\author{M. Constantinou, H. Panagopoulos, A. Skouroupathis}
\address{Department of Physics, University of Cyprus, P.O. Box 20537,
Nicosia CY-1678, Cyprus \\
{\it email: }{\tt phpgmc1@ucy.ac.cy, haris@ucy.ac.cy, php4as01@ucy.ac.cy}}
\vskip 3mm

\date{\today}

\maketitle

\begin{abstract}

We study a systematic improvement of perturbation theory for gauge fields
on the lattice; the improvement entails resumming, to all orders in
the coupling constant, a dominant subclass of tadpole diagrams.

This method, originally proposed for the Wilson gluon
action~\cite{PV1}, is extended here to encompass all possible gluon
actions made of closed Wilson loops; any fermion action can be
employed as well.
The effect of resummation is to replace various parameters in the
action (coupling constant, Symanzik coefficients, clover coefficient)
by ``dressed'' values; the latter are solutions to certain
coupled integral equations, which are easy to solve numerically.

Some positive features of this method are: a) It is gauge invariant, b) it
can be systematically applied to improve (to all orders) results
obtained at any given order in perturbation theory, c) it does indeed
absorb in the dressed parameters the bulk of tadpole contributions.

Two different applications are presented: The additive
renormalization of fermion masses, and the multiplicative renormalization
$Z_V$ ($Z_A$) of the vector (axial) current. In many cases where
non-perturbative estimates of renormalization 
functions are also available for comparison, the agreement with
improved perturbative results is significantly better as
compared to results from bare perturbation theory.

\medskip
{\bf Keywords:} 
Lattice QCD, Perturbation theory, Improved actions, Tadpole improvement.

\end{abstract}

\newpage


\section{Introduction}

Since the earliest studies of quantum field theories on a lattice, it
was recognized that quantities measured through numerical simulation
are characterized by significant renormalization effects, which must be
properly taken into account before meaningful comparisons to
corresponding physical observables can be made.

As has been rigorously demonstrated~\cite{Reisz}, the renormalization
procedure can be formally carried out in a systematic way to any given
order in perturbation theory. However, calculations are notoriously
difficult, as compared to continuum regularization schemes;
furthermore, the convergence rate of the resulting asymptotic series
is often unsatisfactory.

A number of approaches have been pursued in order to improve the
behaviour of perturbation theory, among them
Refs.~\cite{Parisi,L-M}. These approaches share in common the aim
to reorganize perturbative series in terms of an expansion coefficient
which would be more suitable than the bare coupling constant $g_0$;
the definition of such a ``renormalized'' coupling constant is not
unique, but can depend on the observables under study and on an energy
scale. It is expected that such a definition will reabsorb a large
part of the tadpole contibutions which are known to
dominate lattice perturbation theory.

Some years ago, a method was proposed to sum up a whole subclass of
tadpole diagrams, dubbed ``cactus'' diagrams, to all orders in
perturbation theory~\cite{PV1,PV2}; this procedure has a number of
desirable features: It is gauge invariant, it can be systematically
applied to improve (to all orders) results obtained at any given order
in perturbation theory, and it does indeed absorb the bulk of tadpole
contributions into an intricate redefinition of the coupling constant;
in cases where non-perturbative estimates of renormalization
coefficients are also available for comparison, the agreement with
cactus improved perturbative results is significantly better as
compared to results from bare perturbation theory.

In the present work we extend the improved perturbation
theory method of Refs.~\cite{PV1,PV2}, to encompass the large class of
actions which are used 
nowadays in simulations of QCD. This class includes Symanzik improved
gluon actions with any arbitrary combination of closed Wilson loops,
combined with any fermionic action. 
In Section II we present our
calculation, deriving expressions for a dressed gluon propagator, as
well as for dressed gluon and fermion vertices, as a result of the
summation of cactus diagrams to all orders. We show how these dressed
constituents are employed to improve 1-loop and 2-loop Feynman
diagrams coming from bare perturbation theory. In Section III we apply
our improved renormalizaton procedure to a number of test cases
involving Symanzik gluons and Wilson/clover/overlap fermions. 
Improvement of QED is relegated to an Appendix. 

Clearly, all resummation procedures, whether in the continuum or on
the lattice, bear a caveat: A one-sided resummation could
ruin desirable partial cancellations which might exist among those
diagrams which are resummed and others which are not; what is worse,
the end result might depend on the gauge. As we shall see, no partial
cancellations will be ruined in our procedure, due to the distinct
$N$-dependence of the resummed diagrams ($N$ is the number of colors);
furthermore, our results will be gauge independent.

\section{The Method}
In this Section, following the outline of Ref.~\cite{PV1}, we start
illustrating our method by showing how the 
gluon propagator is dressed by the inclusion of cactus
diagrams. We will then dress gluon and fermion vertices as well. 
Finally, we will explain how this procedure is applied to
Feynman diagrams at a given order in bare perturbation theory,
concentrating on the 1- and 2-loop case.

\subsection{Dressing the propagator}
We consider, for the sake of definiteness, the Symanzik
improved gluon action involving Wilson loops with up to 6 links; it
will be evident from what follows that the method is applicable to 
any gluon action made of Wilson loops. In standard notation (see,
e.g., Ref.~\cite{HPRSS}), the action reads: 

\begin{eqnarray}
S_G=\frac{2}{g_0^2} \,\,& \Bigg[ &c_0 \sum_{\rm plaquette} {\rm Re\,
    Tr\,}(1-U_{\rm plaquette})\,  
  +  \, c_1 \sum_{\rm rectangle} {\rm Re \, Tr\,}(1- U_{\rm
    rectangle}) \nonumber \\  
 & + & c_2 \,\sum_{\rm chair} {\rm Re\, Tr\,}(1- U_{\rm chair})
\, 
  +  \, c_3 \sum_{\rm parallelogram} {\rm Re \,Tr\,}(1-
U_{\rm parallelogram})\Bigg]\,
\label{gluonaction}        
\end{eqnarray}
The coefficients $c_i$ can in principle be chosen arbitrarily, subject
to a normalization condition which ensures the correct classical
continuum limit of the action:
\begin{equation}
c_0 + 8 c_1 + 16 c_2 + 8 c_3 = 1 \label{norm}
\end{equation}
Some popular choices of values for $c_i$
used in numerical simulations will be considered in the applications
of Section III. 
The quantities $U_i$ ($i = 0,1,2,3,$ respectively: plaquette,
rectangle, chair, parallelogram) 
in Eq. (\ref{gluonaction}) are products of 
link variables $U_{x,\mu}$ around the perimeter of the closed
loop. 

Applying the usual parameterization of links in terms of the continuum
gauge fields $A_\mu(x)$ 
\begin{equation}
U_{x,\mu} = \exp\,\Bigl(i\,g_0\,a\, A_\mu(x+a\hat\mu/2)\Bigr),
 \qquad A_\mu(x)=
A_\mu^a(x)\, T^a, \quad {\rm Tr}\,(T^a\,T^b) = {\textstyle {1\over
    2}} \,\delta^{ab}
\end{equation}
($a$\,: lattice spacing, set to one from now on; $\hat\mu$\,: unit vector in
direction $\mu$\,; $T^a$\,: generator of $SU(N)$ algebra)
and the Baker-Campbell-Hausdorff (BCH) formula, $U_i$ takes the form:
\begin{equation} 
U_i = \exp\,\Bigl(i\, g_0\, F^{(1)}_i + i\, g_0^2\, F^{(2)}_i +
 i\, g_0^3\,F^{(3)}_i + {\cal O}(g_0^4)\Bigr)
\label{BCH}\end{equation}
where $F^{(1)}_i$ is simply the sum of the gauge fields on the links
of loop $i$ (e.g., for the plaquette: $F^{(1)}_0 = A_\mu(x{+}\hat\mu/2) +
A_\nu(x{+}\hat\mu{+}\hat\nu/2) - A_\mu(x{+}\hat\nu{+}\hat\mu/2) 
- A_\nu(x{+}\hat\nu/2)$), while
$F^{(j)}_i\ (j>1)$ are $j$-th degree polynomials in the gauge
fields, constructed from nested commutators.

\noindent
\begin{minipage}{0.50\linewidth}
\ \ \ We may now {\it define} the cactus diagrams which dress the gluon
propagator: These are tadpole diagrams which become disconnected
if any one of their vertices is removed (see Fig. 1); further, each
vertex is constructed solely from the $F^{(1)}_i$ parts of the action.

\ \ \ A diagrammatic equation for the dressed gluon propagator (thick line)
in terms of the bare propagator (thin line) and 1-particle irreducible (1PI)
vertices (solid circle) reads:
\end{minipage}\hskip0.05\textwidth
\begin{minipage}{0.38\linewidth}
\medskip
{\centerline{\psfig{figure=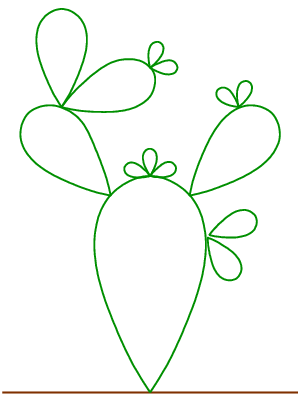,width=3truecm}}}
\medskip
{\centerline{\bf Figure 1: A cactus}}
\end{minipage}

\medskip
\begin{equation}
\psfig{figure=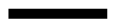,width=1truecm}
= \psfig{figure=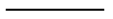,width=1truecm} +
\psfig{figure=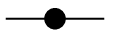,width=1truecm} +
\psfig{figure=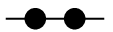,width=1truecm} + \cdots
\label{propdress}
\end{equation}
 The 1PI vertex obeys the following recursive equation:
\begin{equation}
\psfig{figure=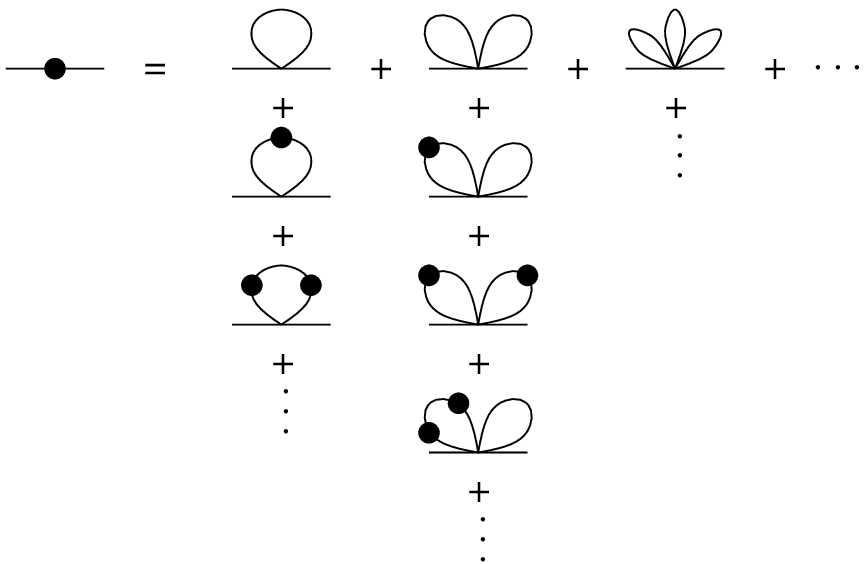,width=10truecm}
\label{recursive}
\end{equation}
In order to put this equation into a mathematical form and solve it,
let us first write down the bare inverse propagator $D^{-1}$ resulting
from the action~(\ref{gluonaction}), and from the gauge fixing term: 
\begin{equation}
S_{gf} = {1\over 1-\xi}\, \sum_{x,\mu , \nu}
\hbox{Tr} \, \Bigl( \Delta^-_\mu A_\mu(x{+}\hat\mu/2) \,\Delta^-_\nu
A_\nu(x{+}\hat\nu/2)\Bigr) , \qquad
\Delta^-_\mu \phi(x) \equiv \phi(x{-}\hat\mu) - \phi(x)
\end{equation}
The quadratic part of the total gluon action thus becomes (in the
notation of Ref.~\cite{HPRSS}):
\begin{equation}
S_{\rm G}^{(0)} = \frac{1}{2}\int_{-\pi}^\pi \frac{d^4k}{(2\pi)^4}
\sum_{\mu\nu} 
A_\mu^a(k)\,D^{-1}_{\mu\nu}(k)\,A_\nu^a(-k)
\end{equation}
where:\qquad \qquad $\displaystyle D^{-1}_{\mu\nu}(k) = \sum_\rho \left(
\hat{k}_\rho^2 \delta_{\mu\nu} - \hat{k}_\mu\hat{k}_\rho \delta_{\rho\nu}
\right)  \, d_{\mu\rho} + {1\over 1-\xi}\, \hat{k}_\mu\hat{k}_\nu$

\noindent
and:\qquad\qquad $\displaystyle d_{\mu\nu}=\left(1-\delta_{\mu\nu}\right)
\left[C_0 -
C_1 \, \hat{k}^2 -  C_2 \,(\hat{k}_\mu^2 + \hat{k}_\nu^2)
\right],\qquad
 \hat{k}_\mu = 2\sin\frac{k_\mu}{2}\,, \quad
        \hat{k}^2 = \sum_\mu \hat{k}_\mu^2 $

\noindent
The coefficients $C_0,\, C_1,\, C_2$ are related to the Symanzik
coefficients $c_i$ by 
\begin{equation}
C_0 = c_0 + 8 c_1 + 16 c_2 + 8 c_3 \,, \,\,\,
C_1 = c_2 + c_3\,, \,\,\, C_2 = c_1 - c_2 - c_3 
\end{equation}
The inverse propagator can thus be put in the form:
\begin{equation}
D^{-1}_{\mu\nu}(k) \equiv c_0\,G^{(0)}_{\mu\nu}(k) 
+ c_1\,G^{(1)}_{\mu\nu}(k)
+ c_2\,G^{(2)}_{\mu\nu}(k) + c_3\,G^{(3)}_{\mu\nu}(k) 
 + {1\over 1-\xi}\, \hat{k}_\mu\hat{k}_\nu
\end{equation}
The matrices $G^{(i)}(k)$ are symmetric and transverse,
i.e. they satisfy:
\begin{equation}
\sum_\nu G^{(i)}_{\mu\nu}(k)\,\hat k_\nu = 0
\end{equation}
Each of them originates from a corresponding term: 
${\rm Tr}\bigl(F^{(1)}_i\,F^{(1)}_i\bigr)$ of the gluon action.
Consequently, each of the diagrams on the r.h.s. of
Eq.~(\ref{recursive}), being the result of a contraction with only two
powers of $F^{(1)}_i$ left uncontracted, will necessarily be equal to a
linear combination of $G^{(i)}(k)$; this implies that the 1PI vertex
$G^{\rm 1PI}(k)$ (the l.h.s. of Eq.~(\ref{recursive})) can be written as:
\begin{equation}
G^{\rm 1PI}(k) = \alpha_0\,G^{(0)}(k) 
+ \alpha_1\,G^{(1)}(k)
+ \alpha_2\,G^{(2)}(k) + \alpha_3\,G^{(3)}(k)
\end{equation}
Each of the quantities $\alpha_i$ will in general depend on $N,\ g_0,
\ c_0,\ c_1,\ c_2,\ c_3$, but not on the momentum. We must now turn
Eq.~(\ref{recursive}) into a set of 4 recursive equations for
$\alpha_i$\,.

Eq.~(\ref{propdress}) leads to the following expression for the
dressed propagator $D^{\rm dr}(k)$:
\begin{eqnarray}
D^{\rm dr} &=& D + D\,G^{\rm 1PI}\,D + D\,G^{\rm 1PI}\,D\,G^{\rm 1PI}\,D
+ \cdots = D\,\Biggl({\openone\over \openone-G^{\rm 1PI}\,D}\Biggr)\\
\Rightarrow\quad (D^{\rm dr})^{-1} &=& 
(\openone - G^{\rm 1PI}\,D)\,D^{-1} = D^{-1} - G^{\rm 1PI}\nonumber\\
&=& \tilde c_0\,G^{(0)} + \tilde c_1\,G^{(1)} +
\tilde c_2\,G^{(2)} +\tilde c_3\,G^{(3)} + 
{1\over 1-\xi}\, \hat{k}_\mu\hat{k}_\nu\,,\qquad 
\tilde c_i\equiv c_i-\alpha_i\label{Ddressed}
\end{eqnarray}

We observe that dressing affects entirely the transverse part of the inverse
propagator, replacing the bare coefficients $c_i$ with improved ones
$\tilde c_i$\,, and leaves the longitudinal part intact. The same property
carries over directly to the propagator itself; the consequence
of this will be that our method leads to the {\it same results in all
covariant gauges}.

In terms of the dressed propagator, Eq.~(\ref{recursive}) can be {\it
  drawn} as:
\begin{equation}
\psfig{figure=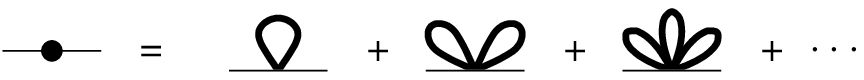,width=10truecm}
\label{recursive2}
\end{equation}
Let us now evaluate a typical diagram on the r.h.s. of
Eq.~(\ref{recursive2}); it will be the sum of 4 terms, one term for each of
the Wilson loops $U_i\ (i=0,1,2,3)$ in the action, from which its
$n$-point vertex may 
have originated. There are $(n-2)/2$ 1-loop integrals in the diagram;
each of them corresponds to the contraction of two powers of
$F^{(1)}_i$ via a dressed propagator, and will contribute one power of
$\beta_i(\tilde c_0, \tilde c_1, \tilde c_2, \tilde c_3)$, where: 
\begin{eqnarray}
\beta_0 &=&\int_{-\pi}^\pi \frac{d^4q}{(2\pi)^4} \,
\Bigl(2\,\hat q_\mu^2\,D^{\rm dr}_{\nu\nu}(q) - 2\, \hat q_\mu\,
\hat q_\nu\,D^{\rm dr}_{\mu\nu}(q)\Bigr)\nonumber\\
\beta_1 &=&\int_{-\pi}^\pi \frac{d^4q}{(2\pi)^4} \,
\Bigl((4\hat q_\nu^2 - \hat q_\nu^4)\,D^{\rm dr}_{\mu\mu}(q)+
      \hat q_\mu^2 (4{-} \hat q_\nu^2)\,D^{\rm dr}_{\nu\nu}(q) 
      -2\,\hat q_\mu\,\hat q_\nu (4{-} \hat q_\nu^2)\,
D^{\rm dr}_{\mu\nu}(q)\Bigr)\nonumber\\
\beta_2 &=&\int_{-\pi}^\pi \frac{d^4q}{(2\pi)^4} \,
\Bigl(\hat q_\mu^2 (8{-} \hat q_\nu^2)\,D^{\rm dr}_{\rho\rho}(q)/2
-\hat q_\mu\,\hat q_\rho (8{-} \hat q_\nu^2)\,
D^{\rm dr}_{\mu\rho}(q)/2\Bigr)
\nonumber\\
\beta_3 &=&\int_{-\pi}^\pi \frac{d^4q}{(2\pi)^4} \,
\Bigl(3\,\hat q_\mu^2 (4{-} \hat q_\nu^2)\,D^{\rm dr}_{\rho\rho}(q)/2
-3\,\hat q_\mu\,\hat q_\nu\, (4{-} \hat q_\rho^2)\,
D^{\rm dr}_{\mu\nu}(q)/2\Bigr)\label{beta}
\end{eqnarray}
($\mu, \nu, \rho$ assume distinct values; no summation implied).
Once again, we note that $\beta_i$ are gauge independent, since the
longitudinal part cancels in the loop contraction.

For the contraction of the $SU(N)$ generators we first
evaluate $F(n;N)$, which is the sum over all complete pairwise
contractions of ${\rm Tr}\{T^{a_1} T^{a_2}\ldots T^{a_n}\}$:
\begin{eqnarray}
F(2;N) &=& \delta_{a_1 a_2}\,{\rm Tr}\left\{T^{a_1}\, T^{a_2} \right\}
\nonumber\\
F(4;N) &=& (\delta_{a_1 a_2}\,\delta_{a_3 a_4}+
\delta_{a_1 a_3}\,\delta_{a_2 a_4}+\delta_{a_1 a_4}\,\delta_{a_2
  a_3})\,{\rm Tr}\left\{T^{a_1}\, T^{a_2}\, T^{a_3}\, T^{a_4} \right\}
\nonumber\\
F(n;N) &=& {1\over 2^{n/2} (n/2)!} \sum_{P\in S_n} \delta_{a_1 a_2}\,
\delta_{a_3 a_4} \ldots \delta_{a_{n-1} a_n}\, {\rm Tr} 
\left\{T^{P(a_1)}\, T^{P(a_2)} \ldots T^{P(a_n)} \right\}
\end{eqnarray}
($F(2n+1;N)\equiv 0$; $S_n$ is the permutation group of $n$ objects).
The generating function $G(z;N)$ for this quantity:
\begin{equation}
G(z;N) \equiv \sum_{n=0}^\infty {z^n\over n!} F(n;N) \qquad 
\Rightarrow \qquad F(n;N) = {d^n\over dz^n} G(z;N)\big|_{z=0}
\label{GzN}
\end{equation}
has been computed explicitly in Ref.~\cite{PV1}, using Gaussian
integration over the space of traceless Hermitian
matrices~\cite{Mehta}, with the result:
\begin{equation}
G(z;N) = e^{z^2 (N-1)/(4N)} \, L^1_{N-1}(-z^2/2)
\label{Gresult}
\end{equation}
($L^\alpha_\beta(x)\,$: Laguerre polynomials). Since 2 out of $n$
generators remain uncontracted in our case, color contraction does not
lead to $F(n;N)$, but rather to:
\begin{equation}
{n \, F(n;N)\over 2( N^2 {-}1)}
\label{contraction2}
\end{equation}
Thus, upon contraction, an $n$-leg diagram in Eq. (\ref{recursive2}),
with its vertex 
coming from the term $U_i$ of the Langrangian ($i=0,1,2,3$), will merely
result in the following multiple of $G^{(i)}$ :
\begin{equation}
{c_i\over g_0^2}\, {(i\,g_0)^n\over n!}\,
{n \, F(n;N)\over 2( N^2 {-}1)}\,4\,\beta_i^{(n-2)/2}\,G^{(i)}
\end{equation}

We are finally in a position to set Eq. (\ref{recursive2}) in a
mathematical form:
\begin{eqnarray}
\alpha_0\,G^{(0)} 
+ \alpha_1\,G^{(1)}
+ \alpha_2\,G^{(2)} + \alpha_3\,G^{(3)} =&& \nonumber\\
\sum_{i=0}^3 \,\sum_{n=4,6,8,\ldots}^\infty &&{c_i\over g_0^2}\,
{(i\,g_0)^n\over n!}\,
{n \, F(n;N)\over 2( N^2 {-}1)}\,4\,
\beta_i^{(n-2)/2}\,G^{(i)}\label{recursive3}
\end{eqnarray}
Unknown in Eq.~(\ref{recursive3}) are the coefficients $\alpha_i$\,;
they appear on the l.h.s., as well as inside the integrals $\beta_i$ of
the r.h.s, by virtue of Eqs. (\ref{beta}, \ref{Ddressed}). 
We recall that $G^{(i)}$ are functions of the external momentum $k$\,;
if these are independent\footnote{Actually, $G^{(2)}$ is not
  independent of the rest, so that we have 3 equations for 3
  coefficients; this causes no complication. In any case, typically
  $c_2=0$ in simulations.}, 
then Eq.~(\ref{recursive3}) amounts to 4
equations for the 4 coefficients $\alpha_i$\,.

The generalization of our procedure for improved gluon actions with
arbitrary numbers and types of Wilson loops is now evident.
It is crucial to check at this stage that all combinatorial weights are
correctly incorporated in Eq.~(\ref{recursive3}); this is indeed the case.

Splitting Eq.~(\ref{recursive3}) into 4 separate equations, and making
use of Eq.~(\ref{GzN}), we can recast the infinite summations
in closed form:
\begin{eqnarray}
{\alpha_i\over c_i} &=& \sum_{n=4,6,8,\ldots}^\infty \,{1\over g_0^2}\,
{(i\,g_0)^n\over n!}\,{n \, F(n;N)\over 2( N^2 {-}1)}\,4\,
\beta_i^{(n-2)/2}\nonumber\\
&=&1+\left(\sum_{n=0}^\infty \,{(i\,g_0)^n\over n!}\, F(n{+}1;N)\,
\beta_i^{n/2}\right) {2(i\,g_0)\over g_0^2\,(N^2{-}1)}\,
\beta_i^{-1/2} \nonumber\\
&=& 1 - {2\over z\,(N^2{-}1)} \,
G^\prime(z;N)\Bigr|_{z=(i\,g_0\,\beta_i^{1/2})}
\label{aOVERc}
\end{eqnarray}
\begin{eqnarray}
\Rightarrow\quad {c_i{-}\alpha_i\over c_i}\,(N^2{-}1) &=&
{2\over z}\,G^\prime(z;N)\Bigr|_{z=(i\,g_0\,\beta_i^{1/2})}
\nonumber\\
&=&e^{-\beta_i\,g_0^2\,(N{-}1)/(4N)}\,\left({N{-}1\over N}\,L^1_{N-1}
(g_0^2\,\beta_i/2) + 2\, L^2_{N-2}(g_0^2\,\beta_i/2)\right)
\label{cimp}
\end{eqnarray}

In solving Eqs.~(\ref{cimp}), each choice of values for 
($c_i\,,\,g_0\,,\,N$)
leads to a set of values  for $\tilde c_i\equiv c_i-\alpha_i$\,. The
latter are no longer normalized in the sense of Eq.~(\ref{norm});
one may equivalently choose, however, to express the results of our
procedure in terms of a normalized set of improved coefficients,
$\tilde c_i/\tilde C_0$ and an improved coupling constant $\tilde
g_0^2 = g_0^2/\tilde C_0$, where: 
$\tilde C_0 = \tilde c_0 + 8 \tilde c_1 + 16 \tilde c_2 + 8 \tilde
c_3$\,.
In fact, it is convenient to treat bare and improved coefficients on
an equal footing, by defining rescaled quantities as
follows\footnote{The dressed propagators in $\tilde\beta_i$
  will now contain a rescaled gauge parameter $(1{-}\xi)\to
  g_0^2\,(1{-}\xi)$, which is irrelevant since the longitudinal
  part does not contribute.}:
\begin{equation}
\gamma_i \equiv {c_i\over g_0^2}\,,\qquad
\tilde\gamma_i \equiv {\tilde c_i\over g_0^2}\,,\qquad
\tilde\beta_i(\tilde c_0, \tilde c_1, \tilde c_2, \tilde c_3) \equiv
g_0^2 \,\beta_i(\tilde c_0, \tilde c_1, \tilde c_2, \tilde c_3) =
\beta_i(\tilde\gamma_0, \tilde\gamma_1, \tilde\gamma_2,
\tilde\gamma_3)
\end{equation}
The rescaled quantities $\tilde\gamma_i$ must now satisfy the
coupled equations:
\begin{equation}
\tilde\gamma_i = {1\over N^2{-}1}\, \gamma_i \,
e^{-\tilde\beta_i\,(N{-}1)/(4N)}\,\left({N{-}1\over N}\,L^1_{N-1}
(\tilde\beta_i/2) + 2\, L^2_{N-2}(\tilde\beta_i/2)\right)
\label{rescaled}\end{equation}
For the gauge groups $SU(2)$ and $SU(3)$, the Laguerre polynomials
have a simple form, making Eqs.~(\ref{rescaled}) more explicit:
\begin{equation}
(N=2)\,:\ \tilde\gamma_i = \gamma_i \,e^{-\tilde\beta_i/8}\,
\left(1-{\tilde\beta_i\over 12}\right)\,,\qquad
(N=3)\,:\ \tilde\gamma_i = \gamma_i \,e^{-\tilde\beta_i/6}\,
\left(1-{\tilde\beta_i\over 4}+{\tilde\beta_i^2\over 96}\right)
\label{N2N3}
\end{equation}
Given the highly nonlinear nature of Eqs.~(\ref{rescaled}), it is not
{\it a priori} clear that a solution for $\tilde\gamma_i$ always
exists\footnote{The converse, of course, is trivial: Finding the bare values
$\gamma_i$ which lead to a given set of dressed values
$\tilde\gamma_i$ is immediate, since the integrals $\tilde\beta_i$ only
depend on $\tilde\gamma_i$, not $\gamma_i$\,.}; it turns out that this
is always the case, for all physically interesting values of $c_i$, and for all
values of $g_0$ ranging from $g_0=0$ up to a certain limit value, well
inside the strong coupling region.

Fortunately, numerical solutions of Eqs.~(\ref{rescaled}, \ref{N2N3})
can be found very easily. We use a fixed point procedure, applicable
to equations of the type $x = f(x)$\,: 
\begin{equation}
\tilde\gamma_i = f_i(\tilde\gamma_i) \qquad\Rightarrow\qquad
\tilde\gamma_i = \lim_{m\to\infty}\,\tilde\gamma_i^{(m)}, \qquad
{\rm where:\ } \tilde\gamma_i^{(0)}=\gamma_i\,,\quad 
\tilde\gamma_i^{(m{+}1)}= f_i(\tilde\gamma_i^{(m)})
\end{equation}
In order for the procedure to converge (attractive fixed point), it
must be that: $|\partial f_i/\partial \tilde\gamma_i| < 1$ in a
neighborhood of $\tilde\gamma_i$. This has
been verified in a number of extreme cases.

\subsection{Numerical values of improved coefficients}
Here we present the values of the
dressed coefficients for several gluon actions of interest. 

\vspace{0.5cm}
\begin{center}
\psfig{figure=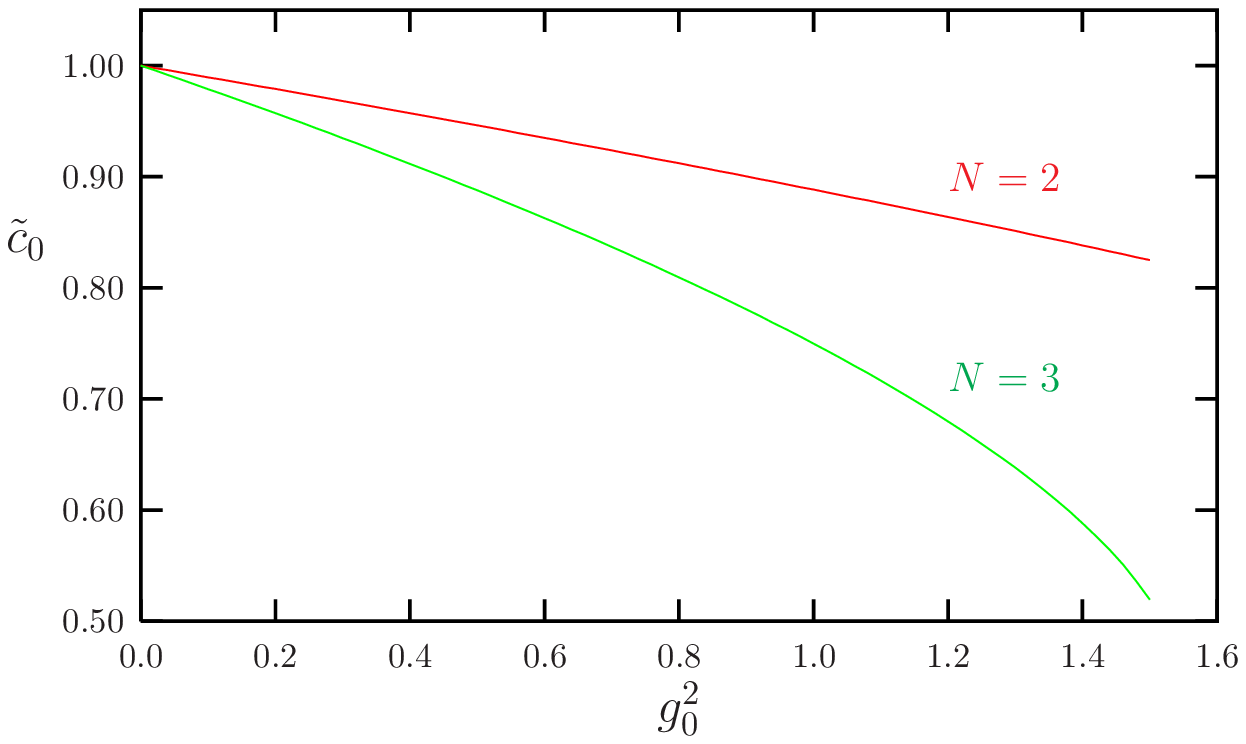,width=10truecm}
\vskip 2mm
\small{\label{fig2}Fig. 2: Improved coefficient $\tilde c_0$ for
  $N{=}2$ and $N{=}3$ (plaquette action)}
\end{center}

\noindent
$\bullet$\ \ Let us start with
the plaquette action ($c_0{=}1$, $c_1{=}c_2{=}c_3{=}0$)~\cite{PV1}: 
In this case,
Eqs.~(\ref{rescaled}) reduce to only one equation, for
$\tilde\gamma_0$, while $\tilde\gamma_i{=}\gamma_i{=}0\ (i{=}1,2,3)$. This
equation is further simplified greatly since the integral
$\tilde\beta_0$ can now be evaluated in closed form,
$\tilde\beta_0 = 1/(2\tilde\gamma_0)$; for $N=3$ we obtain
(cf. Eq.~(\ref{N2N3})): 
\begin{equation}
\tilde c_0 = e^{\displaystyle -g_0^2/(12\,\tilde c_0)}\,
\left(1-{g_0^2\over8\,\tilde c_0}+{g_0^4\over384\,\tilde c_0^2}\right)
\end{equation}
In Fig. 2 we plot $\tilde c_0$ (in the notation of~\cite{PV1},
$\tilde c_0\equiv1{-}w(g_0)$) as a function of $g_0^2$, for $N=2$ and
$N=3$. The range of $g_0$ values, for which solutions exist, 
extends from $g_0^2=0$ (where $\tilde c_0 = 1$) up to $16\sqrt{e}/3
\simeq 3.23$ ($N=2$) and $1.558$ ($N=3$); this covers the whole
region of physical interest.

\vspace{0.5cm}
\begin{center}
\psfig{figure=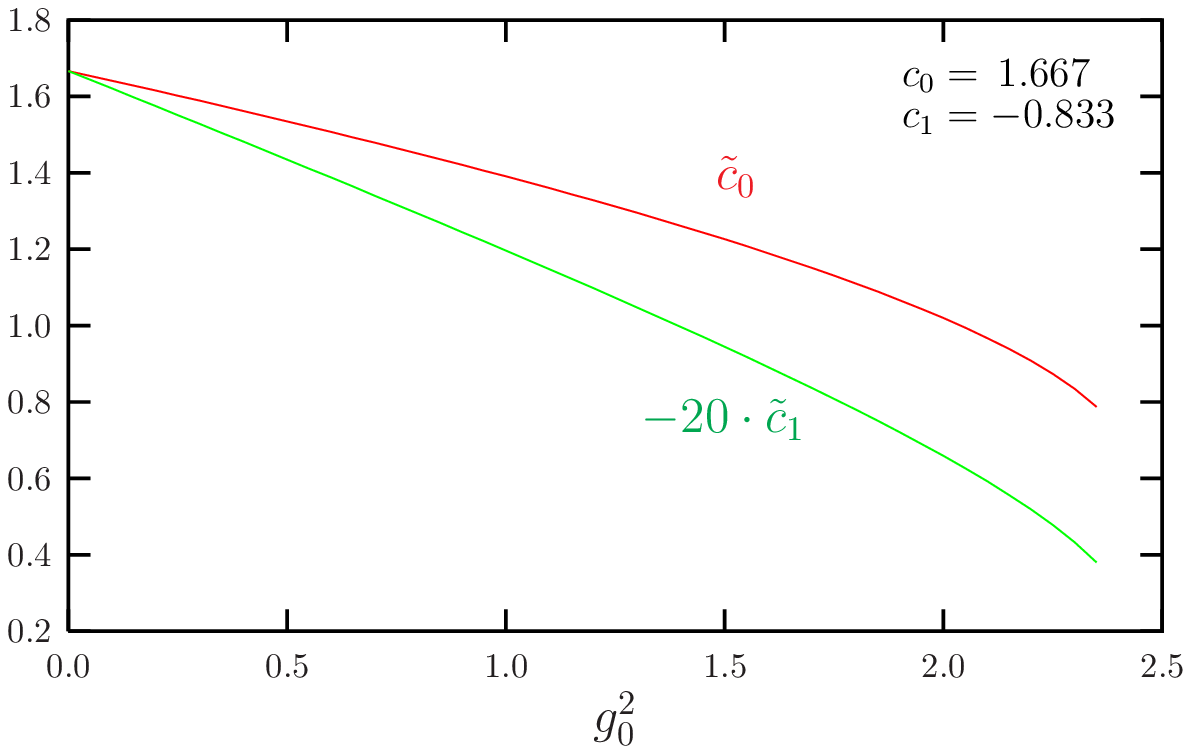,width=11truecm}
\vskip 2mm
\small{\label{fig3}Fig. 3: Improved coefficients $\tilde c_0$ and 
$\tilde c_1$ (tree-level Symanzik improved action)}
\end{center}

\noindent
$\bullet$\ \ The tree-level Symanzik improved action~\cite{Symanzik}
corresponds to: $c_0{=}5/3$, $c_1{=}-1/12$, $c_2{=}c_3{=}0$. The
dressed coefficients $\tilde c_0,\,\tilde c_1$ are shown in Fig. 3 for
$N=3$.

\vspace{0.5cm}
\begin{center}
\psfig{figure=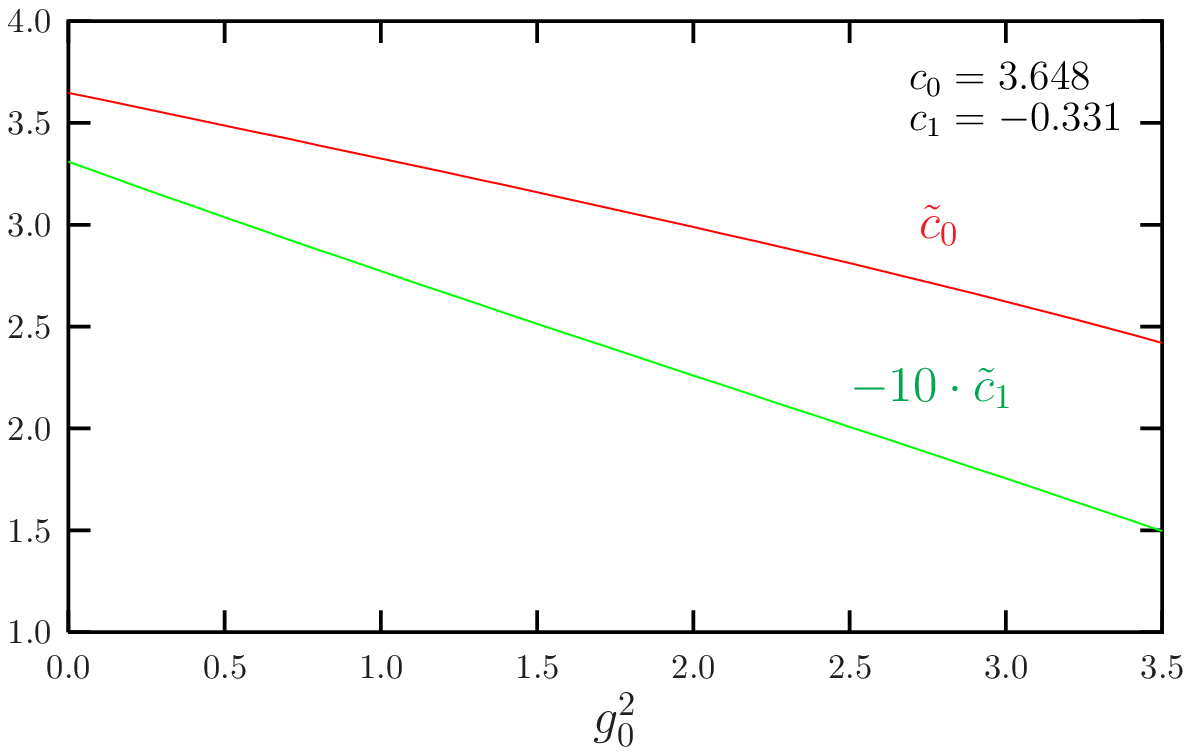,width=11truecm}
\vskip 2mm
\small{\label{fig4}Fig. 4: Improved coefficients $\tilde c_0$ and 
$\tilde c_1$ (Iwasaki action)}
\end{center}

\noindent
$\bullet$\ \ The Iwasaki set of parameter values~\cite{Iwasaki} is: 
$c_0{=}3.648$, $c_1{=}-0.331$, $c_2{=}c_3{=}0$; while, in principle,
$c_0$ and $c_1$ depend on $g_0$, they are typically kept constant in
simulations. The corresponding dressed values are plotted in Fig. 4 ($N=3$).

\vspace{0.5cm}
\begin{center}
\psfig{figure=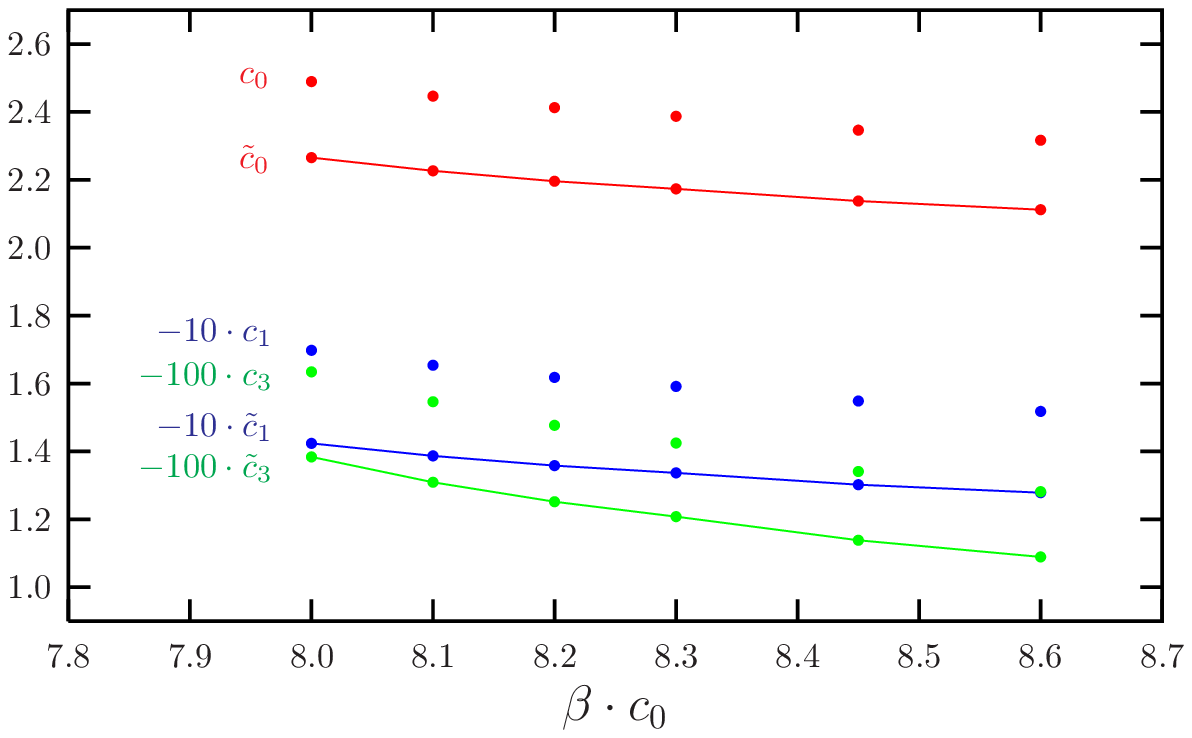,width=11truecm}
\vskip 2mm
\small{\label{fig5}Fig. 5: Coefficients $c_0\,,\,c_1\,,\,c_3$
  (red/blue/green dots, respectively) and their dressed counter\-parts 
$\tilde c_0\,,\,\tilde c_1\,,\,\tilde c_3$ (dots joined by a line), 
for different values of $\beta\,c_0 =
  6\,c_0/g_0^2$ (TILW actions)} 
\end{center}

\noindent
$\bullet$\ \ Another class of gluon actions based on Symanzik
improvement are the tadpole improved L\"uscher-Weisz (TILW)
actions~\cite{Luscher,Alford}. In this case, the coefficients
$c_0\,,\,c_1\,,\,c_3$ are optimized for each value of $\beta = 2N/g_0^2$
separately ($c_2=0$). In Fig. 5 we show the values of $c_i$ and of
their dressed counterparts $\tilde c_i$ in a typical range for
$\beta$\,: $8.0 \le \beta\,c_0 \le 8.6$ ($N=3$).

\medskip
\begin{center}
\begin{minipage}{11cm}
\begin{center}
\begin{tabular}{cr@{}lr@{}lr@{}lr@{}l}
\hline
\multicolumn{1}{c}{$\beta$}&
\multicolumn{2}{c}{$c_0$}&
\multicolumn{2}{c}{$c_1$}&
\multicolumn{2}{c}{$\tilde{c}_0$} &
\multicolumn{2}{c}{$\tilde{c}_1$} \\
\tableline \hline
1.1636  &$\phantom{A^{3^2}}$5&.29078  &$\phantom{000}$-0&.53635  &
$\phantom{000}$3&.39826           &$\phantom{000}$-0&.22528    \\
0.6508  &12&.2688  &-1&.4086   &8&.8070            &-0&.7313     \\
\hline
\end{tabular}

\vskip 3mm
\small{TABLE I. Improved coefficients $\tilde{c}_0$ and $\tilde{c}_1$ in the
  DBW2 action, for $\beta=6/g_0^2=1.1636$ and $0.6508$}
\end{center}
\end{minipage}
\end{center}

\medskip
\noindent
$\bullet$\ \ Finally, the DBW2 gluon action~\cite{Takaishi}
corresponds to $c_2{=}c_3{=}0$, and $\beta$-dependent values for
$c_0$, $c_1$. Some standard values for $c_0$ and $c_1$ (obtained
{\it starting} from $\beta\,c_0=6.0$ and $6.3$), as well as $\tilde{c}_0$ 
and $\tilde{c}_1$ are shown in Table I.

\subsection{Dressing vertices}
\noindent
$\bullet$\ \ We will begin by considering the 3-gluon vertex, coming from the
action, Eq.~(\ref{gluonaction}). This vertex results from a Taylor
expansion of $U_i$ to 3rd order in $g_0$\,. Expressing $U_i$ as in
Eq.~(\ref{BCH}), we see that only terms of the form ${\rm Tr}(F^{(1)}_i\,
F^{(2)}_i)$ will appear in this vertex, since 
${\rm Tr}(F^{(3)}_i)$ and ${\rm Tr}\bigl({(F^{(1)}_i)}^3\bigr)$ will
vanish.

By analogy with Eq.~(\ref{recursive2}), the dressed 3-gluon vertex
equals:
\begin{equation}
\psfig{figure=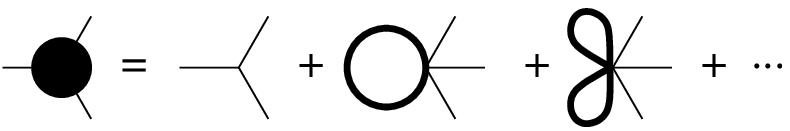,width=8truecm}
\label{threedressed}
\end{equation}
Consistently with the dressing of propagators, each $(2l+1)$-point
vertex in Eq.~(\ref{threedressed}) is a sum of 4 parts (one from each
type of Wilson loop in the action), made up of 
\begin{equation}
{\rm Tr}\bigl({(F^{(1)}_i)}^{2l-1}F^{(2)}_i\bigr)
\end{equation}
Denoting the bare 3-gluon vertex by: $V_3 = c_0\,V_3^{(0)} +
c_1\,V_3^{(1)} + c_2\,V_3^{(2)} + c_3\,V_3^{(3)}$, 
it is relatively
straightforward to see from Eq.~(\ref{threedressed}) that the dressed
vertex, $V_3^{\rm dr}$, is given by:
\begin{equation}
V_3^{\rm dr} = \sum_{i=0}^3 \,c_i\,\left(\sum_{l=0}^\infty 
{(i\,g_0)^{2l{+}1}\over (2l{+}1)!}\,{2\over (N^2{-}1)}\,
F(2l{+}2;N)\,
\beta_i^l\right)\,(i\,g_0)^{-1}\,V_3^{(i)}
\label{3gluon}
\end{equation}
The summations inside parentheses are a mere multiple
of those in Eq.~(\ref{aOVERc}); consequently, the result for 
$V_3^{\rm dr}$ turns out very simple:
\begin{equation}
V_3^{\rm dr} = \tilde c_0\,V_3^{(0)} +
\tilde c_1\,V_3^{(1)} + \tilde c_2\,V_3^{(2)} + 
\tilde c_3\,V_3^{(3)}
\label{3gluon2}
\end{equation}

\smallskip\noindent
$\bullet$\ \ We turn now to the 3-point fermion-antifermion-gluon
vertex~\cite{PV2}. In the cases of Wilson and overlap fermions, these
vertices remain unaffected, since the fermion actions do not contain
any closed Wilson loops on which the BCH formula might be
applied. The vertex from the
clover action, on the other hand, is amenable to improvement; we
write:
\begin{equation}
\psfig{figure=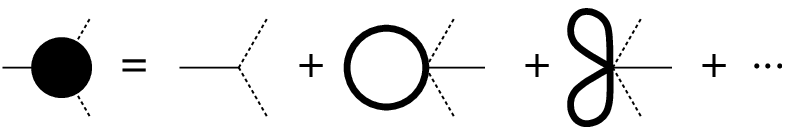,width=8truecm}
\end{equation}
(fermions are denoted by a dotted line). Just as in
Eq.~(\ref{3gluon}), we find:
\begin{equation}
\psfig{figure=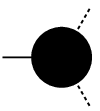,width=1truecm}
\ \ \ {=\atop \phantom{{\Bigl(0\over0}}}
\ \ \ \psfig{figure=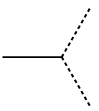,width=1truecm}\ \ 
{\displaystyle\cdot \left(\sum_{l=0}^\infty 
{(i\,g_0)^{2l}\over (2l{+}1)!}\,{2\over (N^2{-}1)}\,
F(2l{+}2;N)\,\beta_0^l\right)\  =\atop \phantom{{0\over0}}}
\ \ \ \psfig{figure=threeFermionBare.eps,width=1truecm}\ \ 
{\displaystyle\cdot\left({\tilde c_0\over c_0}\right)\atop 
\phantom{{0\over0}}}
\label{cswDressed}
\end{equation}

\smallskip\noindent
$\bullet$\ \ Vertices with more fields would seem {\it a priori} more difficult to
handle. To illustrate the complications that may arise, let us
consider the 4-gluon vertex. The BCH expansion of ${\rm Tr}(U_i)$
contributes to this vertex in the form: 
${\rm Tr}\bigl({(F^{(1)}_i)}^4\bigr)$, 
${\rm Tr}\bigl({(F^{(2)}_i)}^2\bigr)$ and 
${\rm Tr}(F^{(1)}_i\,F^{(3)}_i)$. Such terms may in principle dress
differently from each other. In addition, the dressed vertex produced
from ${\rm Tr}\bigl({(F^{(1)}_i)}^4\bigr)$ will not be a multiple of
its bare counterpart; rather, it will be a linear combination of two color
tensors (which are independent for $N>3$): 
\begin{equation}
{\rm Tr} \{ T^a T^b T^c T^d + {\rm permutations}\}\quad{\rm and }\quad
(\delta^{ab} \delta^{cd} + \delta^{ac} \delta^{bd} +
\delta^{ad} \delta^{bc} )
\end{equation}
This issue has been resolved in Ref.~\cite{PV1}, and it generalizes
directly to the present case. Actually, such
complications will not appear while dressing 1- and 2-loop diagrams in
typical cases: Terms of the type ${\rm Tr}\bigl({(F^{(1)}_i)}^4\bigr)$
must simply be omitted in order to avoid double counting, since their
contribution is already included in dressing diagrams with one
less loop. Thus, one is left only with: ${\rm Tr}\bigl({(F^{(2)}_i)}^2\bigr)$ and 
${\rm Tr}(F^{(1)}_i\,F^{(3)}_i)$; for both of these terms it is
straightforward to show, just as in Eqs.~(\ref{3gluon},
\ref{3gluon2}), that their
dressing amounts to replacing $c_i$ by $\tilde c_i$\,.

\smallskip\noindent
$\bullet$\ \ The same considerations as above apply to all higher
vertices from both the gluon and fermion actions as well.

\subsection{The improvement procedure in a nutshell}
The steps involved in the resummation of cactus diagrams can now be
described quite succinctly:
\begin{itemize}
\item Substitute gluon propagators in Feynman diagrams by their
dressed counterparts. The latter are obtained by the replacement $c_i
\to \tilde c_i = g_0^2\,\tilde\gamma_i$\,, where $\tilde \gamma_i$ are
the solutions of Eqs.~(\ref{rescaled}).
\item Perform the same replacement, $c_i\to \tilde c_i$\,, on the 3-gluon
  vertex.
\item Account for dressing of the 3-point vertex from the clover
  action by adjusting the clover coefficient $c_{\rm SW}$\,: 
$c_{\rm SW} \to c_{\rm SW}\cdot(\tilde c_0/c_0)$
\item In dressing subleading-order diagrams, avoid double counting,
  i.e., subtract terms which were included in dressing leading-order 
  diagrams. These are very easy to identify and subtract:
  Writing a general subleading-order result (aside from an overall
  prefactor) as: $a/N^2 + b + c\,N_f/N$ ($N_f$\,: number of fermion
  flavors),
 the quantity to subtract will
  include all of $a/N^2$ (because terms with BCH commutators are higher
  order in $N$), and it will be a multiple of $(2N^2-3)$; thus
  subtraction boils down to the substitution:
\begin{equation}
a/N^2 + b + c\,N_f/N \to \left(b+{2\over3}\,a\right) + c\,N_f/N
\end{equation}
(see Refs.~\cite{PV1,FP,PP,SCP} for different applications of this).
The remaining subleading vertices dress exactly as the
  propagators and 3-point vertices above.

\end{itemize}

\section{Applications}
We turn now to two different
applications of cactus improvement: The additive mass renormalization
for clover fermions and the 1-loop renormalization of the axial and
vector currents using the overlap action. 
Both cases employ Symanzik improved gluons; hence, our
results are presented for various sets of Symanzik coefficients. 

\subsection{Critical mass of clover fermions}
It is well known that an ultra-local discretization of the fermion
action without doubling breaks
chirality. Consequently, we must demand a zero renormalized fermion
mass, in order to ensure chiral symmetry while approaching the continuum
limit. For this purpose, the bare mass is additively renormalized from
its zero tree-level value to a critical value $dm$.

We compute $dm$ using clover fermions and Symanzik improved
gluons, in 1-loop perturbation theory; the result, denoted as 
$dm_{\rm 1-loop}$, is then dressed with cactus diagrams to arrive at
the improved value $dm_{\rm 1-loop}^{\rm dr}$.
The reader can refer to other works \cite{FP,PP} for
more details on the definition of $dm$. A two-loop calculation of $dm$
with the same actions can be found in our Ref.~\cite{SCP}.
There are only two 1-loop
diagrams contributing to $dm_{\rm 1-loop}$ and the result can be
written as a polynomial in the clover parameter:

\begin{eqnarray}
dm_{\rm 1-loop} = \sum_{i=0}^2 {\varepsilon^{(i)} c_{\rm SW}^{i}} 
\nonumber \\  
dm_{\rm 1-loop}^{\rm dr} = \sum_{i=0}^2 {\varepsilon^{(i)}_{ \rm dr} 
c_{\rm SW}^{i}}
\label{dm}        
\end{eqnarray}
Clearly, one-loop results are independent of $N_f$, the number of fermion flavors.
$\varepsilon^{(i)}_{\rm dr}$ includes one factor of $\tilde{c_0}/{c_0}$ for each power of $c_{\rm SW}$ (cf. Eq~(\ref{cswDressed})). 
An overall factor of $g_0^2\,(N^2{-}1)/N$ has been absorbed
in the coefficients $\varepsilon^{(i)}$ and $\varepsilon^{(i)}_{\rm dr}$,
since the improvement procedure requires us to choose definite values of
$g_0$ and $N$.

Some numerical values of Eqs. (\ref{dm}) corresponding to the plaquette
and Iwasaki actions are given below ($N=3$). First, for the
plaquette action, choosing $\beta=6.0$ one gets
\begin{equation}
\begin{array}{lclllll}
dm_{\rm 1-loop}&=&-0.43428549(1)&+0.1159547570(3)&c_{{\rm SW}}&+0.0482553833(1)&c_{{\rm SW}}^2 \\  [0.5ex]
dm_{\rm 1-loop}^{\rm dr}&=&-0.579221119(2)&+0.1159547570(3)&c_{{\rm SW}}&+0.0361806779(1)&c_{{\rm SW}}^2
\end{array}
\label{dmPlaq}
\end{equation}
in agreement with Ref. \cite{PP}. For the Iwasaki action at $\beta=1.95$\,:
\begin{equation}
\begin{array}{lclllll}
dm_{\rm 1-loop}&=&-0.6773690760(3)&+0.2342165224(9)&c_{{\rm SW}}&+0.0806966864(3)&c_{{\rm SW}}^2 \\  [0.5ex]
dm_{\rm 1-loop}^{\rm dr}&=&-0.757856451(1) &+0.1671007819(8)&c_{{\rm SW}}&+0.0447467282(1)&c_{{\rm SW}}^2
\end{array}
\label{dmIw}
\end{equation}

Eqs. (\ref{dmPlaq}, \ref{dmIw}) can be used to evaluate the critical hopping
parameter $\kappa_{\rm cr}$ through:
\begin{equation}
\kappa_{\rm cr}\equiv{1\over 2\,dm + 8\,r} \nonumber
\end{equation}
where $r$ is the Wilson parameter. Estimates of $\kappa_{\rm cr}$ from
numerical simulations exist in the
literature for the plaquette action \cite{Bowler,LSSWW} ($N_f=0$),
\cite{UKQCD,JS} ($N_f=2$), and also the
Iwasaki action \cite{Khan} ($N_f=2$). Perturbative (unimproved and dressed) and
non-perturbative results are listed in Table II for specific
values of $c_{\rm SW}$. It is clear that cactus dressing
leads to results for $\kappa_{\rm cr}$  which are much closer to values
obtained from simulations. 

\medskip
\begin{center}
\begin{minipage}{15cm}
\begin{center}
\begin{tabular}{lclllr@{}llr@{}llr@{}llr@{}l}
\hline
\multicolumn{1}{l}{Action}&$N_f$&\phantom{aa}&
\multicolumn{1}{c}{$\beta$}&$\phantom{\biggl(\biggr)}$&
\multicolumn{2}{c}{$c_{\rm SW}$}&\phantom{aa}&
\multicolumn{2}{c}{$\kappa_{\rm cr, 1-loop}^{\phantom{\rm dr}}$} &\phantom{aa}&
\multicolumn{2}{c}{$\kappa_{\rm cr, 1-loop}^{\rm dr}$} &\phantom{aa}&
\multicolumn{2}{c}{$\kappa_{\rm cr}^{\rm non-pert}$} \\
\tableline \hline
Plaquette\phantom{aa}&0&&6.00&$\phantom{A^{3^2}}$&
                 1&.479        &&0&.1301      &&0&.1362    &&0&.1392\\
Plaquette&0&&6.00&&1&.769      &&0&.1275      &&0&.1337    &&0&.1353\\
Plaquette&2&&5.29&&1&.9192     &&0&.1262      &&0&.1353    &&0&.1373\\
Iwasaki  &2&&1.95&&1&.53       &&0&.1292      &&0&.1388    &&0&.1421\\
\hline
\end{tabular}

\vskip 3mm
\small{TABLE II. 1-loop results and non-perturbative values for
  $\kappa_{\rm cr}$}
\end{center}
\end{minipage}
\end{center}

\subsection{One-loop renormalization of fermionic currents}
As a second application of cactus improvement, we investigate the
renormalization constant $Z_V$ ($Z_A$) of the flavor non-singlet
vector (axial) current in 1-loop 
perturbation theory. Overlap fermions and Symanzik improved gluons
are employed. Bare 1-loop results for $Z_{V,A}$ have been computed in
the literature \cite{AFPV,HPRSS,IP}; they depend on the overlap
parameter $\rho$ ($0<\rho<2$). 

One can show that, using the overlap action, the renormalization
constants $Z_V$ and $Z_A$ are equal \cite{AFPV}; in the
$\overline{MS}$ scheme they read

\begin{equation}
Z_{V,A} (a,p) = 1 - g_0^2 \,z_{1V,A}
\equiv 1 - g_0^2 \,\frac{C_F}{16 \pi^2} \,(b_{V,A}+b_{\Sigma})       
\label{Zv}        
\end{equation}
folowing the notation of~\cite{HPRSS,IP}. $b_{V,A}$ and $b_{\Sigma}$
are 1-loop results pertaining to the amputated two-point function of
the current and to the fermion self-energy, respectively;
since $Z_V=Z_A$, we can write $b_V=b_A$. Using cactus improvement,
Eq. (\ref{Zv}) becomes 
\begin{equation}
Z_{V,A}^{\rm dr} (a,p) = 1 - g_0^2 ~z_{1V,A}^{\rm dr} 
\label{Zvdressed}        
\end{equation}
To compute $z_{1V,A}^{dr}$ we dress
the Symanzik coefficients and the propagators as described in the
previous section. In Table III the values of $Z_{V,A}$ and
$Z_{V,A}^{\rm dr}$ are presented for different sets of the Symanzik
coefficients, choosing $\rho=1.0$, $\rho=1.4$. Systematic errors are too
small to affect any of the digits appearing in the table. The
dependence of $Z_{V,A}$ and $Z_{V,A}^{\rm dr}$ on the
overlap parameter $\rho$ is more clearly shown in Fig. 6, where we 
plot our results for three actions: Plaquette, Iwasaki and
TILW. Note that improvement is more apparent for
the case of the plaquette action. Indeed, from Table III one can
clearly see that the effect of dressing is smaller for
improved gluon actions. This, of course, could have been expected,
since these actions were constructed in a way as to reduce lattice
artifacts, in the first place.

\medskip
\begin{center}
\begin{tabular}{lllclclclc}
\hline
\multicolumn{1}{l}{Action$\phantom{{A^A}^A_{A_A}}$}&
\multicolumn{1}{c}{$\beta{=}6/g_0^2$}&\phantom{a}&
\multicolumn{1}{c}{$Z_{V,A} (\rho{=}1.0)$}&\phantom{a}&
\multicolumn{1}{c}{$Z_{V,A}^{\rm dr} (\rho{=}1.0)$} &\phantom{a}&
\multicolumn{1}{c}{$Z_{V,A} (\rho{=}1.4)$} &\phantom{a}&
\multicolumn{1}{c}{$Z_{V,A}^{\rm dr} (\rho{=}1.4)$} \\
\tableline \hline
Plaquette$\phantom{{A^A}^A}$&6.00            &&1.26427              &&1.35247                &&1.14707            &&1.19615                \\
Symanzik    &5.00            &&1.24502              &&1.29231	             &&1.13574            &&1.16207                \\
Symanzik    &5.07            &&1.24164              &&1.28735	             &&1.13386            &&1.15932                \\
Symanzik    &6.00            &&1.20418              &&1.23484	             &&1.11311            &&1.13019                \\
TILW        &3.7120          &&1.27581              &&1.31941	             &&1.15259            &&1.17690                \\
TILW        &3.6018          &&1.28223              &&1.32764	             &&1.15613            &&1.18146                \\
TILW        &3.4772          &&1.28946              &&1.33677	             &&1.16012            &&1.18651                \\
TILW        &3.3985          &&1.29434              &&1.34298	             &&1.16282            &&1.18995                \\
TILW        &3.3107          &&1.29973              &&1.34972	             &&1.16579            &&1.19369                \\
TILW        &3.2139          &&1.30569              &&1.35705	             &&1.16908            &&1.19774                \\
Iwasaki     &1.95            &&1.39343              &&1.44921	             &&1.21724            &&1.24847                \\
Iwasaki     &2.20            &&1.34872              &&1.38773	             &&1.19256            &&1.21440                \\
Iwasaki     &2.60            &&1.29507              &&1.31940	             &&1.16293            &&1.17656                \\
DBW2        &0.6508          &&1.49631              &&1.45362	             &&1.27543            &&1.25057                \\
\hline
\end{tabular}

\vskip 3mm
\small{TABLE III. Results for $Z_{V,A}, Z_{V,A}^{\rm dr}$
(Eq. (\ref{Zv},\ref{Zvdressed})), using $\rho{=}1.0, \rho{=}1.4$}
\end{center}

A comparison between our improved $Z_{V,A}$ values and some
non-perturbative estimates \cite{GHR}, shows that improvement
moves in the
right direction. Cactus dressing had already been tested using
clover fermions \cite{PV2}, and it turns out to be as good as standard
tadpole improvement \cite{L-M}, but still not
very close to non-perturbative results.

\medskip
\begin{center}
\psfig{file=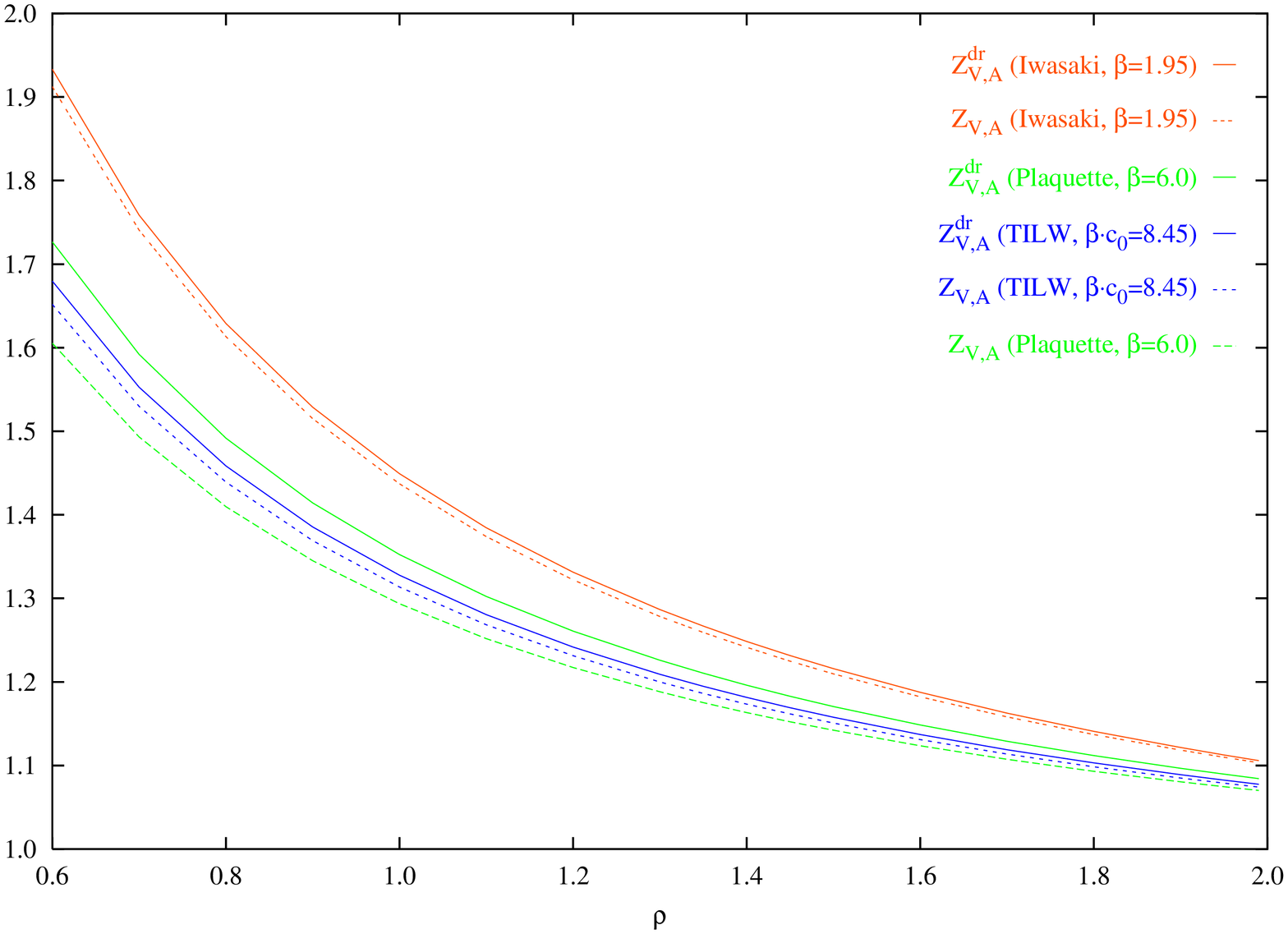,scale=0.6,angle=-90}
\vskip 2mm
\small{Fig. 6: Plots of $Z_{V,A}$ and $Z_{V,A}^{\rm dr}$ for the plaquette,
  Iwasaki and TILW actions. Labels have been placed in the same
  top-to-bottom order as their corresponding curves.}
\end{center}

\vspace*{1.5cm}

\begin{center}
$\phantom{a}^{\star\quad\star\quad\star\quad\star\quad\star 
\quad\star\quad\star\quad\star}$
\end{center}

In closing, we remark that resummation of cactus diagrams is readily
applicable to any observable in lattice gauge theories. This procedure
for improving bare perturbation theory is gauge invariant, and can be
applied in a systematic fashion to 
improve (to all orders) results obtained at any given order in
perturbation theory.

\appendix
\section{Dressing QED}
Cactus improvement can be easily carried over to Lattice Quantum
Electrodynamics. In this case, link variables commute; hence the first
order BCH formula is exact and dressing includes the full contribution
of diagrams with cactus topology.

Dressing the propagator now proceeds precisely as in
Eq.~(\ref{aOVERc}). The only difference is that the result of contracting
$n{-}2$ out of $n$ generators of $SU(N)$: $n \, F(n;N)/(N^2{-}1)$,
must now be replaced by: 
$\displaystyle \left({n\atop 2}\right)\,(n-3)!!$ (the
number of ways to pair $n{-}2$ out of $n$ objects); this results in:
\begin{eqnarray}
{\alpha_i\over c_i} &=& \sum_{n=4,6,8,\ldots}^\infty \,{1\over g_0^2}\,
{(i\,g_0)^n\over n!}\,\left({n\atop 2}\right)\,(n-3)!!\,2\,
\beta_i^{(n-2)/2} = 1 - e^{\displaystyle -\beta_i\,g_0^2/2}\nonumber\\
\Rightarrow \tilde c_i&\equiv& c_i-\alpha_i = c_i\,
e^{\displaystyle -\beta_i\,g_0^2/2}
\label{cimpQED}
\end{eqnarray}
($\beta_i$ are the $\tilde c$-dependent integrals defined in
Eqs.~(\ref{beta})). 

As before, the 4 coupled equations (\ref{cimpQED}) for $\tilde c_i$
assume their simplest form in the case of the plaquette action 
($c_0 = 1,\ c_1=c_2=c_3=0$); in this case, 
$\beta_0 = 1/(2\tilde c_0)$ and we obtain:
\begin{equation}
\tilde c_0 = e^{\displaystyle -g_0^2/(4\tilde c_0)},
\qquad \tilde c_1 = \tilde c_2
= \tilde c_3 = 0
\end{equation}

Dressing vertices is simpler than in the non-Abelian case. For a bare
$(2m)$-point vertex, denoted as: $V_{2m} = c_0\,V_{2m}^{(0)} +
c_1\,V_{2m}^{(1)} + c_2\,V_{2m}^{(2)} + c_3\,V_{2m}^{(3)}$, instead of
contracting group generators, one must simply count the number of
distinct pairings of $2l$ objects out of $(2l{+}2m)$: 
$\displaystyle \left({2l+2m\atop 2m}\right)\,(2l-1)!!$. 
The dressed vertex becomes:
\begin{eqnarray}
V_{2m}^{\rm dr} &=& \sum_{i=0}^3 V_{2m}^{(i)}\,c_i\,\sum_{l=0}^\infty 
\left({(i\,g_0)^{2l{+}2m}\over (2l{+}2m)!}\right)\,
\left({(2m)!\over (i\,g_0)^{2m}}\right)\,
\beta_i^l\,\left({2l+2m\atop 2m}\right)\,(2l-1)!!\nonumber\\
&=& \sum_{i=0}^3 V_{2m}^{(i)}\,c_i\,e^{\displaystyle -\beta_i\,g_0^2/2} 
= \sum_{i=0}^3 V_{2m}^{(i)}\,\tilde c_i
\label{2mgluon}
\end{eqnarray}
The above relations allow us to summarize the dressing procedure for
QED very briefly:

$\bullet$ Replace $c_i$ by $\tilde c_i$ (as given in Eq.~(\ref{cimpQED}))
  throughout 

$\bullet$ Omit all diagrams which contain any cactus subdiagram, to avoid
  double counting

\bigskip\noindent
{\bf Acknowledgements: } This work is supported in part by the
Research Promotion Foundation of Cyprus (Proposal Nr: 
$\rm ENTA\Xi$/0504/11, $\rm ENI\Sigma X$/0505/45).


\end{document}